# Tracer Dispersion in a Multi-compartment Structure

**A. Skvortsov[1], B. Suendermann[2], G. Gamble[2], M. Roberts[1], O. Ilaya[1] and D. Pitaliadda[1]**

[1]Human Performance and Protection Division
Defence Science and Technology Organisation, Victoria 3207, Australia

[2]Marine Platforms Division
Defence Science and Technology Organisation, Victoria 3207, Australia

**Abstract**

The release of a contaminant inside a complex structure was investigated experimentally using a multi-compartment acrylic model submerged in a water tank. Dyed saline solutions were released at point locations and measured at discrete points in the structure using conductivity sensors. Experimental results were then used to validate universal scaling and self-similarity laws of fluid transport processes. The results showed a strong correlation to established scaling and self-similarity laws and validate their use for predicting actual contaminant transport in large structures such as buildings and ships.

**Introduction**

The transport of scalar quantities such as gases, contaminants, particles, and temperature, through a confined environment is of significant interest to a number of science domains covering many practical applications (e.g. underground waste disposal, nuclear reactor leaks, filtration, and indoor dispersal [4-6, 10, 11, 13, 17]). Many powerful analytical and numerical methods have been proposed in the past to enable the development of high-fidelity prediction models such as Computational Fluid Dynamics (CFD) and Lagrangian transport simulations. The application of these methods often relies on some level of expert knowledge or advanced computing facilities to produce tangible results. The reliance of these methods on experts and advanced computing facilities limit their applicability in emergency and operational response.

This study was motivated by the need for a simple (and rigorous) mathematical framework for the risk assessment of engineering structures against specific catastrophic scenarios, such as fire or contaminant transport. To be able to apply a candidate framework in an operational setting, it should include basic mathematical calculations based on the 'first-principles' of the underlying transport phenomena (i.e. mass conservation).

The notion of scaling and self-similarity are consistent with the first-principles approach for predicting the transport and dispersion of tracer materials inside structures [1, 15, 18]. This paper investigates the applicability of simple scaling and self-similarity laws for contaminant transport in complex structures. A water-tank experiment is used to validate the scaling and self-similarity laws.

**Scaling Laws for Confined Dispersion**

The advection of tracer particles can be universally characterised by a scaling law for the mean-square displacement of particles [1, 2, 8, 9, 12, 14]:

$$R \propto A t^\alpha \qquad (1)$$

where $R$ is the size of the particle cloud (or the distance from the source), $t$ is the time elapsed since the particle was released, $A$ and $\alpha$ represent scenario-specific and transport-specific parameters respectively. The value of the dimensional parameter $A$ characterises aspects of the scenario, such as the geometry of the structure and the source conditions [9]; the value of the non-dimensional parameter $\alpha$ is a characteristic of the mechanism of tracer transport. For transport governed by pure advection (ballistic dispersion) with flow velocity $v$, $R \sim vt$, $\alpha = 1$. For purely diffusive movement, the size of the particle cloud $R$ with tracer diffusivity $D$ is given by $R \sim (Dt)^{1/2}$, and $\alpha = \frac{1}{2}$. In the general case of non-uniform flow and movement inside a complex environment (where $v \equiv v(\mathbf{r})$, $D \equiv D(\mathbf{r})$), parameters $A$ and $\alpha$ will have non-universal values. For a localised tracer release associated with a point-source of fluid $v \propto 1/R^2$, and $dR/dt \propto 1/R^2$. The scaling-law becomes $R \propto t^{1/3}$, and $\alpha = 1/3$ (for more details see [3]). Closer to the source, nonlinear effects dominate, and the expansion of the plume is characterised by lower values of $\alpha$ (e.g. $\alpha = 1/6$ in [16]). Consequently $0 \leq \alpha \leq 1$, and $A$ and $\alpha$ should characterise the effective dimensionality of the environment for geometrically constrained structures [12].

It is also useful to define the distribution of tracer particle positions using a Probability Distribution Function (PDF) [9, 14]. The functional form of the PDF for the tracer particles can be deduced by employing arguments of self-similarity based on the scaling-law in Equation (1) (for details see [2, 9, 12]). Consider the self-similar form $\theta(r,t) \propto \Phi(\xi)$, where $\Phi(\cdot)$ is an unknown function, $\xi = r/R(t)$ is the scaled distance (similarity variable) and $R(t)$ is given by Equation (1). Motivated by the solutions of the diffusion equation in open space and assuming a stretched-exponential form, define $\Phi(\xi) = \exp(-\xi^\delta)$ and:

$$\theta(r,t) = N \exp(-\xi^\delta), \qquad (2)$$

where $\delta = const$ is a fitting parameter, and N is a normalisation factor. The standard diffusion spread in 3D space corresponds to $\alpha = \frac{1}{2}$ and $\delta = 2$.

For a continuous tracer source, the system eventually reaches a stationary state and time dependency disappears from Equation (2). The functional form of the PDF in this case can be established by employing the ideas of mass conversation. For a continuous source in open space, the tracer concentration follows a power-law scaling:

$$\theta(r) \propto r^{-\gamma}, \qquad (3)$$

where $\gamma \geq 0$ and is related to the dimensionality of the environment $d_e$, and $\gamma = d_e - 1$. In the 3D case, $d_e = 3$, and $\gamma = 2$, and for the 1D case (e.g. pipe-flow), $d_e = 1$, and $\gamma = 0$. It is reasonable to conclude based on the previous discussions, that in an environment with complex morphology, such as a multi-compartment structure, the power-law in Equation (3) holds but with some unknown effective dimensionality $d_e$ in the range of $1 \leq d_e \leq 3$. In such an approach, $\gamma$ is a fitting parameter of the model in the range of $0 \leq \gamma \leq 2$ [3]. In the remainder of this paper, the experimental procedure and results used to investigate and validate these scaling-laws is presented.

**Experimental Method**

The scaling-laws presented in the previous section are investigated experimentally using a water-tank experiment. Figure 1 shows the configuration of the experimental facility used to validate the scaling-laws. A scale-model of a building was constructed and submerged in a water-tank to investigate the transport and dispersion of tracer in a complex environment. Coloured saline solution with known density was used to represent the tracer. Advection was provided along the length of the water-tank by water-pumps located on the upstream and downstream edge. The transport and dispersion of the tracer was then measured using conductivity sensors located at discrete locations inside the building. A set of collinear sensors were positioned along the centre line of each compartment to measure the tracer concentration at various altitudes.

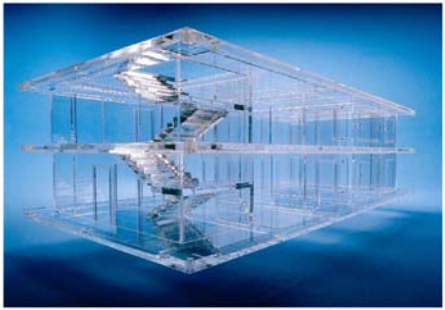

Figure 1. Model of multi-compartment structure used in the water-tank experiments.

The influence of salt concentration, volumetric flow-rate, and morphology of the compartments on the measured concentration was investigated. Salt concentrations for the saline solutions were varied to alter the density of the tracer and investigate the effect of buoyancy on tracer movement. Salt concentrations of $s$ = 1.0%, 2.0%, and 3.0% were used for all experiments. The volumetric flow rates and corresponding Péclet numbers for a tracer with diffusivity $D = 5.53 \times 10^{-6}$ cm$^2$/s [7] are given in Table 1.

| Volumetric Flow Rate ($q$) [mL/s] | Péclet Number ($Pe$) |
|---|---|
| 2.8 | $1.3 \times 10^5$ |
| 1.9 | $8.9 \times 10^4$ |
| 0.95 | $4.4 \times 10^4$ |

Table 1. Volumetric flow rates and associated Péclet numbers used for this experimental study.

Investigation of the effect of morphology on the tracer advection and diffusion was reduced to three limiting configurations:

- **Closed-type, 'C'** – this configuration corresponds to the maximum number of doors closed and only a single path exists connecting any two compartments;
- **Loop-type, 'L'** – this configuration corresponds to the case where a small number of non-unique paths exist between some compartments;
- **Open-type, 'O'** – this corresponds to the maximum number of open doors and the maximum number of loops in the structure's connectivity.

Note, for all cases considered, a continuous path exists connecting two compartments in the structure.

### Results and Discussion

Example sensor response to detected tracer (measured by voltage output) is shown in Figure 2. From Figure 2, the arrival time to each sensor and the saturation limit of in-compartment concentration can be seen. The estimation of these parameters in individual compartments, combined with the geometrical locations of the sensors was used to validate the scaling-laws in Equation (1) and PDFs in Equation (2) and Equation (3).

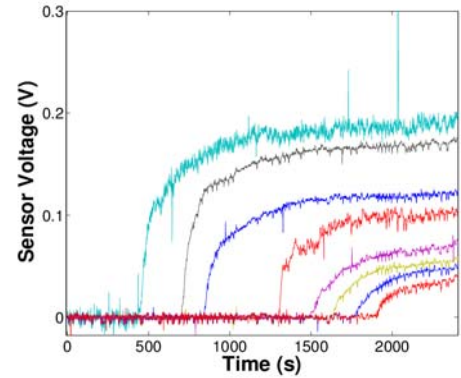

Figure 2. Example signals from individual sensors.

For consistency, it was assumed that the shortest path from the centre of the doorway of each compartment was traversed by the tracer plume (as measured from the plume front). This was used to calculate the transport distances within the structure. Tracer arrival at a sensor was registered when the measured voltage exceeded the threshold level above the noise of the data acquisition system. This corresponded to a threshold level of 0.004 V. Figures 4 -7 shows the distance travelled by the tracer plume to each sensor from the release point as a function of time elapsed from tracer release.

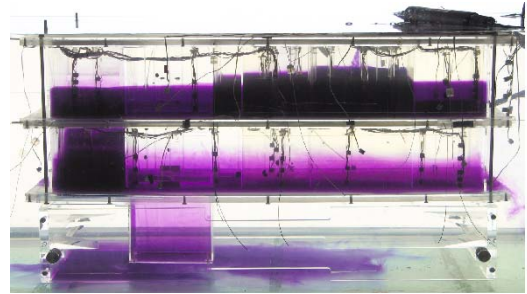

Figure 3. Example of plume propagation through the structure demonstrating characteristic tongue-like propagation associated to the gravity effects.

Results for the first set of experiments are shown in Figure 4. Here, salt concentration $s$ was varied for a fixed flow-rate ($q$ = 95mL) and configuration (C-type). Increasing salt concentration reduces the effective speed of plume propagation and the elapsed time $t$ increases for a given distance $R$ from the source. This can be attributed to the profound contributions of density which is discussed below.

To validate the scaling-laws, it is convenient to rewrite Equation (1) into its non-dimensional form $(R/l_0) \propto A(t/t_0)^{\alpha}$, where $l_0$, $t_0$ are the scale length and time respectively. For the following study, $l_0$ = 1 cm, $t_0$ = 1 s. The power-law dependency makes translation of these results to all other scales straightforward. In fact, for the new scales defined as $l_0 \rightarrow al_0$, $t_0 \rightarrow bt_0$, where $a,b = const$, it only involves rescaling of $A$ in accordance with $A \rightarrow cA$, and $c = a/b^{\alpha}$ (where $\alpha$ is kept the same). Following the results in Figures 4 – 7, values for $A$ and $\alpha$ were determined by plotting the data on

a log-log scale and evaluating the slopes. The estimations of $A$

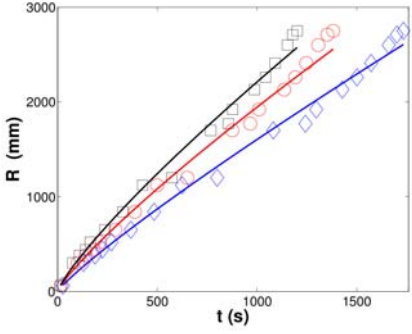

Figure 4. Effect of salt concentration on plume propagation: ◊ - $s$ = 1%, O - $s$ = 2%, □ - $s$ = 3%. Solid lines represent the scaling law for values of the fitted parameters in Table 2.

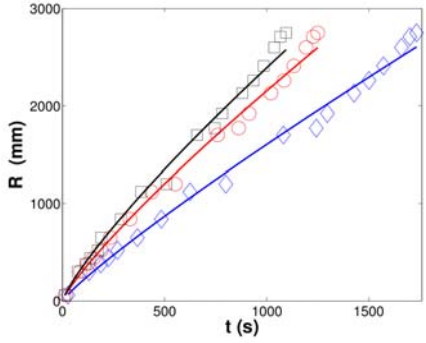

Figure 5. Effect of flow rate on plume propagation: ◊ - $q$ = 0.95 mL/s, O - $q$ = 2mL/s, □ - $q$ = 3mL/s. Solid lines represent the scaling law for values of the fitted parameters in Table 2.

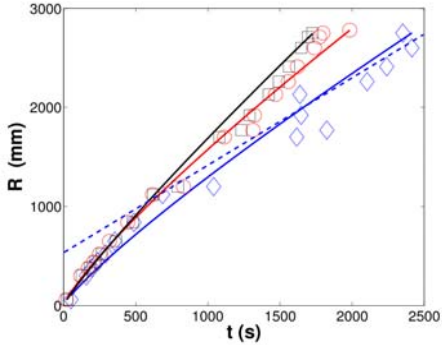

Figure 6. Effect of morphology on plume: ◊ - O-type, O - L-type, □ - C-type. Solid lines represent the scaling law for values of the fitted parameters in Table 2; dashed lines represent new asymptotes for plume propagation. These provide a better fit at later stages for the open morphology.

and $\alpha$ from the set of measurements in Figure 4 are shown in Table 2. Denote $\omega_A(s)$ and $\omega_\alpha(s)$ the non-dimensional sensitivity parameters of $A$ and $\alpha$ for the model with respect to the salt concentration for the measured range respectively; estimates for $\omega_A(s)$ and $\omega_\alpha(s)$ given values $A$ and $\alpha$ from Table 2 are then given by $\omega_A(s) \equiv \langle dA/ds \rangle / \langle A \rangle \approx 0.323$, $\omega_\alpha(s) \equiv \langle d\alpha/ds \rangle / \langle \alpha \rangle \approx -0.03$. Parameters $\omega_A(s)$ and $\omega_\alpha(s)$ are used to calculate the 'shift' of parameters from their reference values, induced by changing salt concentration, i.e. $A \approx A_0[1 + \omega_A(s)(s - s_0)]$, where $A_0$ is the reference value of $A$ for $s = s_0$.

A slight decrease in $\alpha$ emerging from this dataset was consistent with the theoretical predictions for diffusive transport. The increase of salt concentration makes the effect of buoyancy driven transport more profound. This results in the characteristic 'tongue-like' structures shown in Figure 3 that drives the parameter $\alpha$ to its lowest value of 1/3 associated with this process [16]. It is also worth noting that the contribution of the advection mechanism in the overall tracer dispersion diminishes with an increase of the distance from the source, since flow velocity rapidly decays.

| Marker | Salt Variable (Figure 4) | Flow Rate Variable (Figure 5) |
|---|---|---|
| ◊ | $A$ = 3.61 ± 0.09, $\alpha$ = 0.88 ± 0.03 | $A$ = 3.61 ± 0.08, $\alpha$ = 0.88 ± 0.03 |
| O | $A$ = 5.65 ± 0.09, $\alpha$ = 0.85 ± 0.03 | $A$ = 6.41 ± 0.09, $\alpha$ = 0.84 ± 0.03 |
| □ | $A$ = 7.06 ± 0.11, $\alpha$ = 0.83 ± 0.04 | $A$ = 8.01 ± 0.11, $\alpha$ = 0.88 ± 0.03 |

Table 2. Stretched-exponential parameters for varied salt concentrations and flow rate experiments in Figure 4 and Figure 5 respectively.

A possible explanation for some scatter of data points can be drawn from the fact that the arrival times of the tracer plume at the compartments are implicitly dependent not only on the relative distance between compartments, but also on the compartment size (i.e. floor area); i.e. the flow has to back-fill the volume of the compartment to about 1cm to overcome the step into the next compartment. This produces a distribution of 'delays' in the apparent arrival time in the next compartment.

Figure 5 shows the data for the second set of experiments involving the C-type configuration with fixed salt concentration ($s$ = 1%) and varied pump flow rate. Figure 5 shows that by increasing the pump flow rate, the propagation rate of the plume is also increased. This is consistent with the theoretical predictions discussed previously (i.e. increasing pump flow rate increases the relative contribution of advective transport mechanisms) and supported by the monotonic increase in $A$ and $\alpha$ shown in Table 2. Estimates of model sensitivity from $A$ and $\alpha$ with respect to the emitted rate $q$ of tracer source are given by $\omega_A(q) \equiv \langle dA/dq \rangle / \langle A \rangle \approx 0.378$, $\omega_\alpha(q) \equiv \langle d\alpha/dq \rangle / \langle \alpha \rangle \approx -0.04$.

The effect of morphology on arrival times is shown in Figure 6. In general, the morphology of the structure has a twofold influence on the transport properties. Firstly, an increase in structure connectivity potentially reduces the travelling path of the plume as multiple paths are presented. The significant increase in scattered data in the L-type morphology is a direct consequence of the effect of multiple paths leading to each sensor. Formally, the process of faster plume propagation can be captured by increasing parameters $A$ and $\alpha$ in the scaling law in Equation (1). This would provide a better fit for the later stage of plume propagation (dashed line in Figure 6). Secondly, for a given tracer source, the increased connectivity of the structure reduces the effective propagation velocity of the plume and delay transition to the advection-dominated transport mechanism; thus reducing $\alpha$. The reduction of the effective velocity of plume propagation is caused by the associated decrease in the advection velocity of the induced flow inside the structure due to the number of openings. The effect of structure morphology can be considered as an interplay of these concurrent processes and becomes more pronounced with an increase in the distance from the tracer source (i.e. where smaller changes in $A$ and $\alpha$ are more noticeable).

Figure 7 shows experimental data for the concentration distribution inside the structure as a function of the distance from the source and the predictions of the model in Equation (2). Data points in Figure 7 correspond to the experimental settings from Figure 4 and Figure 5. Maximum values for $\theta$ were used as the normalisation factor N in Equation (2). The measurement for the maximum value of $\theta$ corresponds to the concentration of the tracer measured by the sensor closest to the source.

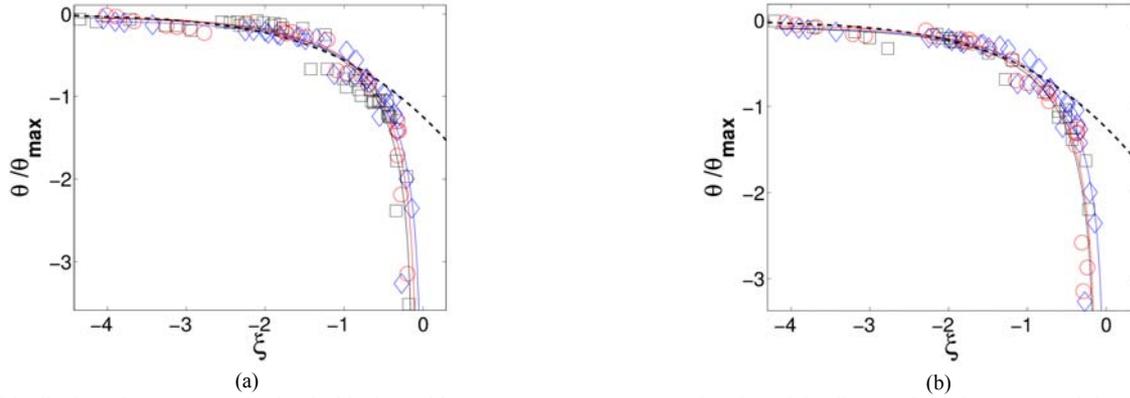

Figure 7. Distribution of tracer concentration inside the multi-compartment structure as a function of the distance from the source and time since release in log-log scale for a) variable salt concentration, and b) variable emission rate of the source. Corresponding experimental data is provided in Figure 4 and Figure 5 respecitvely. Solid lines represent the fitted stretch-exponential profiles from Equation (2) and the dashed lines represent the power-law fit from Equation (4).

From Figure 7, the model in Equation (2) provides a reasonable fit to the experimental data.

Motivated by Equation (3), the following power-law ansatz is proposed to validate the experimental data:

$$\theta(r,t) = \frac{\theta_{\max}}{(1+\xi)^{\gamma}} \qquad (4)$$

where $\gamma = 1.8$ and has been chosen from the 'validity' range $0 \leq \theta \leq 2$. This provides the best fit for the low and intermediate concentrations. It was found that the stretched-exponential distribution in Equation (2) provides a better description of the experimental data (over most of the five orders of magnitude range), while the power-law profile in Equation (4) does not accurately reproduce the shape of the concentration distribution resulting in a poor data fit (at least during the timescale of observations). The fitted values for $\delta$ are: $\Diamond - \delta = 0.95$; $O - \delta = 1.08$; $\square - \delta = 0.92$ for Figure 7a and $\Diamond - \delta = 0.95$; $O - \delta = 1.06$; $\square - \delta = 1.10$ for Figure 7b. It is worth noting that the estimated values of parameter $\delta$ are close to unity and quite consistent.

It can be seen that at a large distance from the source, the distribution given in Equation (3) decays much slower than the distribution in Equation (2). It is worth mentioning that for many practical situations the time taken to reach the saturation limit in Equation (3) can be extremely long, so the transition to the power-law distribution in Equation (3) maybe impractical to observe (see Figure 7).

## Conclusions

Presented are the results of a water-tank experiment of tracer dispersion in a complex multi-compartment structure. The experimental results were shown to have reasonable agreement with a simple theoretical model based on scaling and self-similarity of the underlying transport processes. It is anticipated that the proposed framework can be used to provide a rigorous way to up-scale the results of a laboratory measurement to real-world applications. These can be used as an important step in the development of risk-assessment models for first responders dealing with hazardous releases inside buildings.

## Acknowledgments

We thank I. Burch, A. Douglas and J. Robinson for their help in the setup of the experiment, and S. Cannon, K. Gaylor and C. Woodruff for valuable discussion and support.